\documentclass{eceasst-arxiv}

\usepackage{orcidlink}

\usepackage{color}
\usepackage{fancyvrb}
\usepackage{listings}
\usepackage{nowidow}
\usepackage{subfig}
\usepackage{url}

\usepackage[accsupp]{axessibility}
\input{pygments.tex.in}

\newcommand{\authorRef}[1]{\texorpdfstring{\autref{#1}}{}}
\newcommand{\authorOrcid}[1]{\texorpdfstring{\thinspace\orcidlink{#1}\thinspace}{}}

\newcommand{\eg}{e.\,g.}

\newenvironment{aiuse}{%
  \par\vspace{11pt}%
  \noindent\textbf{Declaration on the use of AI:}\enspace}{%
  \par\vspace{11pt}}

\newenvironment{data}{%
  \par\vspace{11pt}%
  \noindent\textbf{Data availability:}\enspace}{%
  \par\vspace{11pt}}

\title{RSE of a Quantum Transport Code and its Effects}
\author{%
	Christoph Conrads\authorOrcid{0000-0003-3887-9201}\authorRef{1}\authorRef{*},
	Edoardo Di Napoli\authorOrcid{0000-0001-5821-5897}\authorRef{2}}
\institute{%
	\autlabel{1} \email{c.conrads@fz-juelich.de}\\%
	\autlabel{2} \email{e.di.napoli@fz-juelich.de}\\%
	\autlabel{*} Corresponding author\\%
	Jülich Supercomputing Centre, Forschungszentrum Jülich, Germany}
\abstract{
	This paper presents our research software engineering (RSE) experiences
	over two years with libNEGF, a quantum transport code. We describe
	practical approaches to code quality assurance--including continuous
	integration, automated testing, and compiler warning correction--and
	performance engineering through continuous benchmarking. Our systematic
	application of these practices revealed critical defects: uninitialized
	memory reads, out-of-bounds writes, and notably, a misunderstood
	mathematical model in our boundary condition handling. We also document how
	continuous benchmarking exposed performance regressions caused by HPC
	system configuration changes. Our findings provide data points suggesting
	that a dangerous class of defects--equivalent to undefined behavior in
	C/C++ and processor-dependent behavior in Fortran--is as prevalent in
	Fortran scientific codes as elsewhere.
	While libNEGF is implemented in Fortran, most recommendations are
	applicable to scientific software regardless of implementation language,
	and they can be implemented selectively or in their entirety for both new
	and existing projects.
}

\usepackage{listings}

\begin{document}
\maketitle

\section{Introduction}

Within computational physics, density functional theory (DFT) is a model for
examining the electronic structure of many-body systems. The density
functional-based tight binding (DFTB) approximation of DFT is among the most
popular approaches for simulation codes. libNEGF is an implementation of the
non-equilibrium Green's function (NEGF) method and enables DFTB-based
simulations to compute quantities like the electron density or the Hartree
potential \cite{PecchiaPS2008}. The NEGF ansatz delivers sufficiently accurate
results for large numbers of atoms for which, \eg, ab-initio methods are too
expensive in the foreseeable future. libNEGF is implemented in Fortran with GPU
code written in CUDA.

Under the umbrella of the EoCoE-III
project,\footnote{\url{https://www.eurohpc-ju.europa.eu/research-innovation/our-projects/eocoe-iii_en}}
co-development with Forschungszentrum Jülich started in 2024 with the goal of
achieving exascale performance. This article describes the research software
engineering (RSE) efforts of this project and the outcomes. The general aim of
this document is to provide other RSE practitioners with data points to improve
their decision making and prioritization efforts.

The first part of the article concerns software engineering. We start by
presenting the mindset with which we approach the development tasks before
discussing a variety of topics culminating in the introduction of continuous
integration (CI). With CI in place, the code can be built automatically and
static as well as dynamic code analysis can be run at the push of a button. In
the second part, we describe our setup for job submission and continuous
benchmarking. This section requires that the code can be built automatically;
if CI is not available, then code is more likely to fail to build on target
systems. Next, we talk about four highlights of our RSE efforts including
tracking server-side performance changes of an HPC system and a diagnosed bug.
We close with a section about best practices that we should have implemented.

We do not expect readers of this article to be able or willing to implement all
of the suggested measures for a variety of good reasons. We propose instead to
read this paper to get an impression of the problems that are likely to be
present in a code base, how they affect the correctness and performance, and to
which degree these problems can be resolved. Then recommendations can be
applied selectively as the reader sees fit.

In terms of minimum viability, we recommend to set up CI first with the
reader's scientific software at least being successfully compiled in a
container in the case of compiled languages or with the software successfully
running in a container in the case of interpreted languages. With CI set up, a
single gateway for quality assurance is in place with many options to extend
it. For large legacy code bases, finding out how to reliably build the code in
a container together with all of its dependencies may be a time-consuming task
in itself. The outcome is a reusable and up-to-date description of the build
and its dependencies accessible to all project members.

The article is focused on CI and CB with a focus on compiled languages. Readers
need to know how to write and fix code, how to build their code and how to fix
broken builds, how to write tests, and how to use the revision control system
of their choice.

\section{Software Engineering}
\label{sec:software-engineering}

In this section we will focus on software engineering (SE) aspects. For the
purposes of this section, this is anything but performance engineering. We will
explain which steps we took to turn libNEGF into a software that can be
deployed with confidence at the push of a button on a variety of target
systems.

The mindset underlying our SE efforts is explained first before we dive into
code and code history challenges that have to be resolved. Then, the code
development is moved onto a code forge (\eg, GitLab or GitHub) where a CI
pipeline ensures a consistent quality: the code formatting is checked, the code
is built and tested in a variety of configurations. For completeness, we want
to mention that CI is possible without a code forge as well.

Finally, we discuss the beneficial effects of fixing compiler warnings.

\subsection{Mindset}

We assume our simulations to contain a special class of defects called
\emph{untrapped errors} which we introduce in first. Readers will quickly note
that this concept originates from programming language design and may not apply
to the language that they are using. We ask readers to bear with us because in
the following subsection we explain how mixing executable code from different
sources (\eg, through the use of libraries) can cause untrapped errors even in
safe languages (see below for the meaning of ``safe'').

\subsubsection{Untrapped Errors caused by Code Defects}

The mindset underlying our RSE activities is fundamental skepticism
\cite[\S2.2]{OberkampfR2010} applied to a code written in an unsafe language
\cite[p.~2]{Cardelli2004}:
\begin{quotation}
	It is useful to distinguish between two kinds of execution errors: the ones
	that cause the computation to stop immediately, and the ones that go
	unnoticed (for a while) and later cause arbitrary behavior. The former are
	called trapped errors, whereas the latter are untrapped errors.

	A program fragment is safe if it does not cause untrapped errors to occur.
	Languages where all program fragments are safe are called safe languages.
	Therefore, safe languages rule out the most insidious form of execution
	errors: the ones that may go unnoticed.
\end{quotation}
Examples of untrapped errors are reads of uninitialized variables,
out-of-bounds memory accesses or dereferencing an invalid pointer. Examples of
unsafe languages are C, C++, and Fortran independently of the specific standard
used. In C and C++ terms, untrapped errors are a cause of \emph{undefined
behavior}; in Fortran they are a cause of \emph{processor-dependent} behavior.

Untrapped errors are the cause of many well-documented real-world software
problems \cite{Lattner2011,Regehr2010} including many security vulnerabilities.
Possible effects relevant to scientific computing include crashes or silent
corruption of results. These effects \emph{can} occur but they do not
necessarily happen. They may be triggered when changing the compiler, the
compiler version, the compiler's optimization level, something else in the
software stack, or the machine on which the code is compiled. For the sake of
clarity, we provide real-world examples of untrapped errors here.

We dealt once with a C++ program that was not terminating on x86
CPUs whereas on ARMv7 CPUs, the same code would reliably terminate but with an
incorrect output. The cause was an uninitialized
variable.\footnote{\url{https://gitlab.inria.fr/melissa/melissa-sa/-/commit/dc1b1263d69d2dc81d54faec460fac97c441baa4}}
Another example of untrapped errors arose when two tests written in Fortran and
C, respectively, were added to the libNEGF test suite. The two tests were run
successfully in the CI environment in six different configurations on an x86-64
CPU with GCC (see Section~\ref{sec:ci} for details), they ran successfully on
the supercomputer JUWELS Booster when building with the Intel compiler suite
(2021 release) and GCC but the Intel compilers 2023 release correctly diagnosed
a deallocation of unallocated memory. To verify the presence of a problem in
the code, the code was rebuilt with GCC and the undefined behavior sanitizer
enabled.\footnote{The appropriate choice would have been the address
sanitizer.} The sanitizer did not signal problems but the tests suddenly failed
for numerical reasons. These are two typical examples showing how untrapped
errors are present in real-world scientific codes, they have no clear set of
symptoms associated with them, and they may break the program at any time.


The analysis tool named \emph{Stack} found 40\,\% of all Debian packages to
contain at least one instance of untrapped errors \cite{WangZK2013} and another
study found 16\,\% of the Debian packages under consideration to contain
undefined integer behavior \cite{DietzLR2015}. Coverity's static analysis tool
arrived at an average defect density of 0.45 per 1,000 lines of code
\cite{Coverity2011}. These references are not meant to be comprehensive but
just to show how pervasive this problem is. These numbers are emphatically
\emph{lower bounds}. Proving program correctness (and with it, proving the
absence of untrapped errors) is in general not possible. Thus, these studies
focus on a particular set of problems with simplifying assumptions. When
focussing on scientific software, we do not believe these numbers to improve
significantly. As a supporting data point we mention here the static analysis
of CERN code resulting in 40,000 bug fixes in a code base of ca. 3.5 million
lines of C++ code \cite{Clarke2011}. Similar but less recent examples exist for
Fortran, too \cite[\S4.5]{OberkampfR2010}.

Given the statistics above and our fundamental skepticism, we always assume
that our software contains undetected defects. This assumption quickly turned
out to be justified. To date, our quality assurance efforts detected
signed integer overflows,
several out-of-bounds writes,
three cases of reads of uninitialized memory,
a memory leak,
double frees (deallocation of unallocated pointers), and
dereferencing of null pointers.

\subsubsection{Untrapped Errors caused by Build Defects}

All data in a computer are represented by zeros and ones in computer memory.
Meaning is assigned through the operations applied to the data: bytes processed
by the CPU's instruction decoder are taken to be CPU instructions, bytes
processed by the floating-point unit are taken to be floats, and bytes consumed
by a read (load) instruction are taken to be a reference, and so on \cite[\S2,
\S3]{BryantH2011}.

The use of different programming languages within one project is a common
practice in scientific computing \cite{BasiliCC2008}; one of the most common
examples is Python developers calling LAPACK which is written in a
C-derivative, Fortran, and/or assembly. Programs written in different languages
can always interact through the use of files but for performance reasons, it is
much more common to have code from different sources run as part of the same
program. If this is done, then all codes must agree on the representation and
semantics of the data structures in memory, that is, they must all implement
the same \emph{application binary interface} (ABI) and this is not trivially
fulfilled. For example, on 64-bit systems many compiled languages and
interpreters use unsigned 64-bit integers for indexing (\eg, C, C++, and Rust)
whereas others use signed 32-bit integers (\eg, Fortran). Sometimes ABI
compatibility is broken (intentionally or accidentally) by compiler
vendors\footnote{Example: \url{https://gcc.gnu.org/wiki/Cxx11AbiCompatibility}}
and different versions of a library may induce ABI changes.

On modern HPC systems, usually several compilers are available together with
several libraries built in multiple configurations for each compiler. These
library versions may not match the library version of the Linux distribution in
use. 
If a build mixes incompatible libraries or if code is built with one instance
of a library but then uses a different one at run-time, then the combined code
is again broken. That is, if one has two defect-free pieces of code in
different languages and they share data at the binary level, then the combined
software may contain defects that are untrapped errors. Therefore, untrapped
errors are a risk for all HPC users.

\subsubsection{The Mindset in Practice}

In practice, a social challenge is likely to arise from this mindset. Most
contributors in HPC have no formal training in software development and in our
experience in some programming communities, the concept of untrapped errors is
unheard of. Imagine that such a contributor is told that his software contains
many undetected defects. The developer may counter the defect claims saying
that that he fixed all of the bugs that he is aware of. Since there are no
unresolved bugs in the issue tracker and since the simulation output is as
expected, the software must be defect-free for all practical purposes. In this
situation, it is necessary to explain the concept of untrapped errors and
elaborate in detail on its possible consequences. The bugs that were fixed
during development are only those untrapped errors that were triggered on the
developers' machines.

In our experience, previous test efforts may provide only weak support for
claims of correctness. In the case of libNEGF, there exists an extended test
suite but the output of these tests is just compared against output computed by
earlier libNEGF releases and a comparison with other data sources, \eg,
measurements, may simply not be feasible. The original authors of a simulation
software should not be blamed for this shortcoming because the very reason for
the existence of simulation codes is often that conducting experiments is
infeasible or undesirable \cite[\S1.1]{OberkampfR2010}. In our case, the
simulation of material properties is a research topic just as much as the
synthesis of these materials is \cite{MagginiF2021}.

The bottom line is that we suggest projects with unsafe languages to adopt a
mindset that the code contains defects and to make all contributors aware of
the possible consequences of untrapped errors. A large number of the bugs found
in libNEGF since the project's inception are an outcome of adopting this
mindset. Users of a safe language relying on code in different programming
languages should be aware that their software stack may introduce untrapped
errors.

\subsection{Transition to a Code Forge}
\label{sec:code-forge}

At the start of the EoCoE-III project, the libNEGF source code had already been
under version control for more than a decade and a GitHub project existed for
libNEGF. Nevertheless, development efforts were not being organized using the
code forge and consequently, decision making and the origin of data are
sometimes not traceable. As an example, one of the key design aspects of the
GPU code is its manual, synchronous memory management. The original design
envisioned here a way to fully utilize GPU memory but the motivation for this
decision is evident neither from the source code nor from the GitHub issues.

To ensure the survival of such project-specific knowledge, the following
development workflow was implemented: every new feature and bug is first
discussed in an issue, then a feature branch and an associated pull request
(\emph{merge request} in GitLab lingo) are opened, and finally new code is
merged only after assuring its quality (\eg, by reviewing it).

In practice, there is a social aspect when using code forges. Contributors may
habitually use, \eg, e-mail or their employer-internal chat system to discuss
issues. These discussion should be moved to the code forge because they may not
be accessible to all or future contributors and because these discussions may
be hard to find again if needed.

\subsection{Maintaining the Git History}
\label{sec:git-history}

For the purposes of this article, the \emph{default} branch of a repository is
a long-living branch where code changes are accumulated between two releases. A
\emph{feature} branch is an ephemeral branch opened in response to a merge
(pull) request and at the end of its lifetime, it is merged into the default
branch.

git is often the RCS of choice and among other things, git sets itself apart
from other RCS through its ability to handle a history where a commit may have
more than one immediate successor and more than one immediate predecessor in a
single branch. Without constraints on the development process, this can make it
hard to trace the evolution of a piece of code. In one instance, the authors
were engaged in a project were several developers had worked in parallel on
their own feature branch on an HPC system over the course of several weeks and
repeatedly merged code from the feature branches of their coworkers as needed.
This led to a highly complex development history where it was very hard to make sense of the evolution of the code.





Given the small number of project contributors, it was decided to disallow
merging into feature branches with the goal to facilitate project management
and enable traceability. In hindsight, the scheme suffered two problems: it was
never automatically enforced and it did not specify how to handle the immediate
need for a fix present in one feature branch in another feature branch. Since
this regulation was never automatically enforced, some contributors would
create their feature branch from the feature branch with the patch or
contributors would rebase. Reviewers of merge requests would not necessarily
check the git history but only the commit messages. Nevertheless, libNEGF code
development is traceable thanks to the disciplined use of merge requests as
laid out in Section~\ref{sec:code-forge}.

\begin{figure}
	\subfloat[An example history]{%
		\label{fig:git-history}
		\makebox[0.225\textwidth][c]{\includegraphics{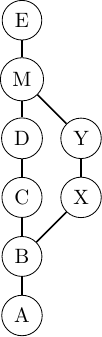}}
	}
	\hfill
	\subfloat[Merge conflict resolution by merging default into feature
	branch]{%
		\label{fig:git-history-merge-conflict-resolution-by-merge}
		\makebox[0.225\textwidth][c]{\includegraphics{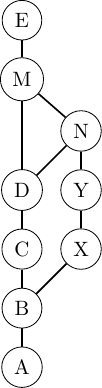}}
	}
	\hfill
	\subfloat[Linear history (rebase followed by fast-forward merge)]{%
		\label{fig:git-history-linear}
		\makebox[0.225\textwidth][c]{\includegraphics{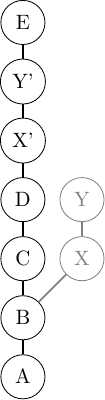}}
	}
	\hfill
	\subfloat[Semi-linear history (rebase followed by merge without fast
	forwarding)]{%
		\label{fig:git-history-semilinear}
		\makebox[0.225\textwidth][c]{\includegraphics{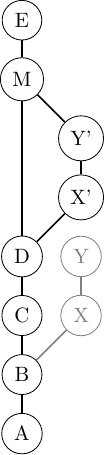}}
	}

	\caption{%
		An overview of the tools to structure a git history. The default branch
		is on the left and circles indicate commits with descendants above
		them. X' and Y' are modified variants of the commit X and Y,
		respectively, that were rebased. Commits labelled with M and N are
		merge commits. The strategies above can be combined with squashing,
		where X and Y (or X', Y', respectively) are combined into a single
		commit.
	}
	\label{fig:git-histories}
\end{figure}

Our first recommendation is to automate enforcement of any branching and
merging strategy employed and educate developers about merge conflict
resolution. Both
GitHub\footnote{\url{https://docs.github.com/en/repositories/configuring-branches-and-merges-in-your-repository/configuring-pull-request-merges/about-merge-methods-on-github}}
and
GitLab\footnote{\url{https://docs.gitlab.com/user/project/merge_requests/methods/}}
support predefined rules for merge requests which are based on the git
primitives of rebasing, squashing, and merging with and without
fast-forwarding. The primitives are illustrated in
Fig.~\ref{fig:git-histories}. The need to inform contributors about merge
strategies arises from the multitude of ways merge conflicts can be resolved in
git because developers may not be aware of all the techniques. In the case of
libNEGF, some contributors would occasionally merge the default branch into the
feature branch as in
Figure~\ref{fig:git-history-merge-conflict-resolution-by-merge}. From the point
of view of a programmer habitually using rebases, this looked like an accident
(with the default branch being mistaken for the feature branch and vice versa)
but it was actually a merge conflict resolution. From the point of view of a
programmer habitually using merges, the interdiction to merge into feature
branches would block their preferred way of merge conflict resolution.

Our second recommendation is to avoid completely unstructured histories to
ensure traceability and facilitate project management, \eg, by forbidding
merging feature branches into feature branches. When failures occur, we are
usually quickly able to pin down the defective commit introducing the problem
because of our merge strategy. With the defective commit at hand, it becomes
possible to revert the commit or to better assess the impact of fixes.

Third, when deciding on a merge strategy we suggest to contemplate the workflow
associated with the different approaches to merge conflict resolution. The
downside of rebasing is that conflicts have to be resolved for every commit in
the feature branch\footnote{\lstinline{git rerere} may be of help:
\url{https://git-scm.com/book/en/v2/Git-Tools-Rerere}} in the worst case. On
the upside, one can automatically build and test the software for every rebased
commit which can help track down errors if the the combined feature and default
branch are not operational. For a merge-based strategy, these points are
reversed: conflict resolution is necessary at most once in the merge commit but
the cause of a non-operational merged branch is harder to find.

\subsection{Small and Fast Tests}

Software testing can be performed at different granularities, \eg, at the level
of individual functions, at the level of individual modules, or at the
application level. Originally, libNEGF possessed only whole-device tests
(\emph{device} here means a quantum device); these are practically system tests
\cite[\S6]{Duvall2007} \cite[\S4.3.3.1]{OberkampfR2010}. Writing tests for
scientific software is often challenging already when the goal is just to test
the solution of simple subproblems, \eg, because there may not exist analytical
solutions or the set of inputs for which an algorithm converges is unknown.

Our suggestion therefore is to have test cases with a limited resource
consumption suited for desktop PCs. This allows contributors to run these tests
frequently on their machines and avoids the need to spend precious compute
hours for this purpose. Furthermore, fast-running tests with a low memory
footprint are amenable to augmentation, i.\,e., additional, expensive checks
can be enabled, the code can be subjected to dynamic analyses, or it can be
debugged on the developer's office machine. With CI, small tests provide faster
feedback to committers.

\subsection{Nonzero Exit on Failure}


The successful execution of a computer program is indicated by its exit status
and it has to be indicated reliably because other parts of the software stack
rely on this information (cf. \lstinline{_Exit()} in
\cite{POSIX}\footnote{\url{https://pubs.opengroup.org/onlinepubs/9799919799/functions/_Exit.html}})
including shell scripts, CI pipelines, and batch schedulers. To the best of our
knowledge, Fortran programs still commonly rely on the STOP intrinsic when
encountering an unrecoverable error. The correct choice in such situations are
the use of \lstinline{STOP x}, where x is a nonzero integer or a string, or
\lstinline{ERROR STOP} if it is available.\footnote{\lstinline{ERROR STOP}
first appeared in the Fortran 2008 standard, cf.
\url{https://fortranwiki.org/fortran/show/Fortran+2008}.}

libNEGF was made to use \lstinline{ERROR STOP} in the case of errors which
immediately lead to the discovery of a broken test. The test in question was
trying to open a nonexisting file, printed an error message to standard output,
and exited with status 0 indicating success.

Our advice for other developers is to
\begin{itemize}
	\item Ensure familiarity of every code contributor with exit codes and
		their common interpretation (zero indicates success, nonzero indicates
		failure).
	\item Determine which tools are offered by a programming language to
		indicate the exit status (\eg, an uncaught Python exception triggers
		automatically a nonzero exit).
	\item Check existing code for exits with status zero in case of error and
		fix these.
\end{itemize}

\subsection{Checking Return Values}
\label{sec:checking-return-values}

We strongly recommend to \emph{always} check \emph{all} return values even in
production runs. In libNEGF, the GPU code was originally discarding return
values and once we started to check them, we immediately discovered several
bugs and found the causes of incorrect simulation outputs.
In one instance an algorithm reliably failed to converge in our tests but this
problem went unnoticed because the iteration count was not checked, see
Section~\ref{sec:broken-math} for details.

Our recommendation is to always check return values. If a return value
indicates failure, there exist a variety of handling strategies
\cite[\S8.3]{McConnell2004} with one approach being to print an error message
and to terminate the simulation with a nonzero status. We suggest this practice
because log files may be cluttered with warnings even during normal execution
and because we do not expect teams to have a process in place to systematically
check the log output for problems when the simulation exits successfully.

Checking the return value may be superfluous when a developer knows that the
return value will \emph{always} indicate success. This is likely never the case
when a function has an error return. For example, MPI is widely used in HPC and
it will never return errors if the the default error handler
\lstinline{MPI_ERRORS_ARE_FATAL} is in use but this error handler can be
changed by every piece of code having access to the associated MPI
communicator. Similarly, \lstinline{malloc()} is unlikely to return NULL
pointers because many operating system are configured to allow overcommitment
of memory. Unless the kernel settings are under your control, there is no
guarantee of this behavior.

\subsection{Continuous Integration}
\label{sec:ci}


We use continuous integration (CI, \cite{Duvall2007}) with great success in
combination with containers \cite{ContainersDocker,McCarthy2018}. The CI setup
itself is simple with a style check and a build followed by a run of the test
suite. We use the ability of our CI provider to run parameterized jobs, i.\,e.,
building and testing is executed several times but with different variable
values; these variables are the target device (i.\,e., whether the code runs on
CPU or GPU), the Linux distribution, and the MPI implementation, see
Table~\ref{tab:build-configurations}. Figure~\ref{fig:libnegf-ci} shows an
abridged version of our GitLab CI configuration and due to space constraints we
cannot elaborate in detail on this file. The key here is the limited overhead
needed to build and test the code in a variety of environments; no changes have
to be made to the simulation code.

\begin{table}
	\begin{center}
			\begin{tabular}{ll}
					\hline
					Linux distribution & Rocky Linux 9, Debian 12 \\
					MPI implementation & MPICH, Open MPI \\
					Compute device & CPU, GPU (CUDA)
			\end{tabular}
	\end{center}
	\caption{Build configurations in use. GPU code is tested only on Debian
	leading to six different configurations overall.}
	\label{tab:build-configurations}
\end{table}

\begin{figure}
	\input{gitlab-ci-libnegf-646d3f8d.yml.tex.in}
	\caption{The CI configuration of libNEGF. The
	file was edited to fit onto one page: the C, C++ compiler flags, the code
	formatting checks, and some CMake arguments were removed. The original file
	can be found at
	\url{https://github.com/libnegf/libnegf/blob/master/.gitlab-ci.yml}}
	\label{fig:libnegf-ci}
\end{figure}

The goal of using different container images is the imitation of different
environments: Rocky Linux is a near-clone of RHEL (which powers most HPC
systems) whereas Debian and its derivatives are frequently found on consumer
machines. MPI is supposed to be implementation agnostic but we test this with
MPICH and OpenMPI. The way we use containers is decidedly different from the
way containers are commonly used: instead of shipping the built software as a
container image to avoid portability problems, we use a multitude of different
containers throughout development to ensure portability. For developers who
prefer to test changes before pushing, the container images can be built and
run locally on a workstation, providing the same environment used by the CI
pipeline.

With this setup we found a bug in the libNEGF C bindings when OpenMPI was in
use. Furthermore, we never faced build problems on Jülich Supercomputing Centre
systems whose configuration closely matches the container setup. On other HPC
systems, the closer the system configuration to our container setup, the
smaller the number of build problems encountered.

We recommend to set up CI as soon as possible with at least one build
environment in which the code can be successfully run. We also suggest to use
container images with all dependencies pre-installed to speed up the CI for
otherwise, it is necessary to download all required packages whenever a
pipeline is launched. To ensure the presence of all dependencies, there may
exist suitable images online (\eg, on DockerHub) and if not, you can build your
own images.

A \emph{Dockerfile} is a set of rules to for building a container and for
simplicity, we show our Debian Dockerfile in Figure~\ref{fig:dockerfile} below.
Clearly, this is a simple file because all libNEGF dependencies can be met with
packages available in the Debian repositories. Sometimes users have to build
dependencies on their own, either manually or with the aid of a package
manager. In this situation Dockerfiles can still be used through
\emph{multi-stage
builds}.\footnote{\url{https://docs.docker.com/build/building/multi-stage/} }
Due to space constraints, we cannot elaborate on this concept here. Dockerfiles
should be checked into version control \cite[pp. 109]{Duvall2007}. Similarly,
while Docker was a container pioneer, many alternatives (\eg, Podman) with
compatible command line interfaces exist nowadays that can consume Dockerfiles.
The Dockerfile in Figure~\ref{fig:dockerfile} can be built and run with
\begin{lstlisting}[language=sh]
docker build --build-arg mpi=openmpi --build-arg device=cpu \
  --tag libnegf-debian -f Dockerfile .
docker run -i -t libnegf-debian
\end{lstlisting}

\begin{figure}
	\input{Dockerfile.tex.in}
	\caption{The Dockerfile for Debian containers (the file was slightly edited
	to fit onto one page). Depending on the values of the build arguments, four
	different container images can be built from one file.}
	\label{fig:dockerfile}
\end{figure}

\subsection{Fix Compiler Warnings}



Coming back to untrapped errors, fixing compiler warnings is considered a best
practice in C and C++, \eg, \cite{McConnell2004} repeatedly mentions paying
attention to compiler warnings and even increasing the warning level. Similar
advice can be easily found on internet forums or in the guidelines of various
industry consortia, \eg, the C and C++ hardening
guidelines\footnote{\url{https://best.openssf.org/Compiler-Hardening-Guides/Compiler-Options-Hardening-Guide-for-C-and-C++}}
of the Open Source Security Foundation. With respect to the actual outcome, the
support for the value of compiler warnings is mostly anecdotal
\cite{KudrjavetsKN2022}. Given our positive experiences with C++ compiler
warnings, the project went ahead and fixed compiler warnings.

Given the comparatively small amount of CUDA code (less than 1,000 lines), a
comprehensive set of warnings was enabled with the flags
\lstinline{-Wextra -Wall -pedantic} in the CI setup. For Fortran, warnings were
enabled selectively because of the large number of warnings generated by the
gfortran \lstinline{-Wall} option. As a compromise, we settled on enabling all
warnings and disabling a subset of warnings considered less relevant for
correctness, \eg, warnings about unused functions. These warnings helped us
find at least two defects in the code including an iteration counter that was
never updated.

Our advice is to increase the warning level and to fix all compiler warnings
immediately: developers should not get used to ignore warnings in their
compiler output; warnings arising from new code should be instantly
recognizable. Since developers may not check the logs of the CI runs and
instead rely only on the success indicator in the GUI out of convenience (this
is not a bad thing), warnings should be turned into errors in the CI.
Obviously, this forces contributors to merge only warning-free code. It is
therefore necessary to discuss with the project members which warnings are the
most relevant and should never be triggered by the code base.

In large code bases, an overwhelming number of warnings may be emitted by flags
like \lstinline{-Wall}. In this situation we suggest to enable a subset of
warnings that is the most likely to uncover bugs and with an amount of warnings
that can be quickly fixed. For example, warnings about uninitialized variables
could be prioritized over warnings of floating-point comparisons with zero. If
you do not know which warnings are most relevant, then you could prioritize by
the number of warnings emitted but this relies on a feature of modern
compilers: they often show the warning together with the flag causing the
compiler to emit this warning. With this feature, you can count the number of
warnings triggered by each compiler flag and then fix the least frequent
warnings first.

Finally, we only encountered one project where all warnings had been disabled.
For instance, the C compiler tried to warn about comparisons that will always
evaluate to the same boolean value, confounding pointers and arrays thereof,
pointers mistaken for integer arguments, and missing declarations (missing
declarations imply no type
checking).\footnote{\url{https://gitlab.inria.fr/melissa/melissa-sa/-/commit/680e58b9982134bb60caa749e37f630256a83f45}}
All of these were true positives and for this reason, we ask readers not to
disable warnings.

\section{Performance Engineering}
\label{sec:performance-engineering}

libNEGF automatically runs strong scaling benchmarks on a weekly basis. The
setup requires an automated benchmarking tool, a GitLab instance, and access to
a supercomputer from the GitLab instance. We want to emphasize that this
section requires the code to build successfully whenever the benchmark is run.
Our CI setup from the previous section ensures that this is the case.

\subsection{Use of a Benchmarking Tool}
\label{sec:jube}

Supercomputers are only accessible by a small set of users and compute time on
them is precious. For this reason it is of utmost importance to know
\emph{what} is being run: the simulation code, its dependencies, the build
options, the simulation input and its outputs including error messages are all
affecting the outcome of a job. For this purpose, a benchmarking tool is very
helpful. This project uses JUBE \cite{BreuerWSMG2024} which downloads the
libNEGF source code, loads the necessary modules, compiles the code, sets up
the simulation, and runs it, and --most importantly-- it performs each run in a
dedicated directory.

This approach provides several major advantages:
\begin{itemize}
	\item It is more convenient to run simulations for scientists.
	\item There is a single point of reference on how to run on a given
		supercomputer.
	\item All of the inputs and outputs of every simulation run are saved.
\end{itemize}
In short, the user can focus on the actual experiment and its outcome.

Regarding the single point of reference, the Jülich Supercomputing Centre is
home to four HPC systems to which the authors have (or had) access. For each
system, one can run on CPU and GPU entailing different build and job scheduler
options. The JUBE script hides the details of these eight configurations.

Note that we follow a strictly layered approach in libNEGF development: it is
still possible to run libNEGF without JUBE and this is occasionally done for
development purposes. In our experience if a certain user needs to deviate from
the setup realized by JUBE, it is often more convenient to modify the JUBE
script. Saving all of the inputs and output provides significant benefit when a
simulation unexpectedly fails because it provides developers with certainty
about the build configuration as well as inputs and outputs of the simulation.

Our recommendation for other developers is simple: use a tool to manage and
archive supercomputer runs.

\subsection{Continuous Benchmarking}

Performance is a key aspect of high-\emph{performance} computing. Continuous
benchmarking (CB) makes performance assessments an integral step of the
development process on the same level as building, testing, and deploying
\cite[\S1]{AltLP2024}. To ensure fulfilling the EoCoE-III performance goals,
the developer team set up CB in mid-2025 with the goal of regular execution on
one of the HPC systems targeted by EoCoE-III (here: JUWELS Booster). Ideally,
we would run CB whenever a contributor pushes code but the compute time on HPC
systems is limited and such an approach would be a significant drain on the
compute time budget.\footnote{One has to apply for access to HPC machines and
we question, too, if the funding agency would agree to a project proposal
spending most of its resources on CB.} Therefore we limit our CB run to once
per week or less frequently.

The CB setup benefits significantly from on-going CB efforts of JSC
\cite{BrommelFS2024,HertenAA2024,BadwaikBR2026}. Before elaborating on the
details of the CB setup, let us briefly review GitLab's CI pipelines. A CI
script describes one or more jobs (not to be mistaken with batch scheduler
jobs) possibly with dependencies between them that are executed whenever a
pipeline is triggered. Each job is executed on a dedicated machine called
\emph{runner}\footnote{\url{https://docs.gitlab.com/runner/}}. Pipelines can
commonly be triggered by pushing to a repository or by a scheduled
event\footnote{\url{https://docs.gitlab.com/ci/pipelines/schedules/}}.

libNEGF stores its benchmarking data together with its JUBE script (see
Section~\ref{sec:jube}) in its own git repository called
\emph{libnegf-benchmarks} in the same GitLab group. For CB, the CI script in
Figure~\ref{fig:libnegf-benchmarks-ci} was added to libnegf-benchmarks and a
schedule was set up. With its choice of tags in lines 29--32, the script
ensures that the GitLab server picks JUWELS Booster as runner. The shell script
in lines 6 to 16 is then executed on a JUWELS Booster front-end node from which
we can submit batch scheduler jobs. Here we re-use the JUBE script from
Section~\ref{sec:jube}; the benchmark to be executed is configured by means of
the environment variables in lines 22--27. The JUBE output is stored in the
persistent storage of the libNEGF compute time project on JUWELS Booster
(persistent storage in contrast to the scratch storage which is purged after
90~days). The CB runs are evaluated on demand with a Python script. In addition
to the JUBE data, the batch scheduler is queried for job information.

Coming back to the CB efforts at JSC, our setup is simple because JSC provides
GitLab runners executing on JUWELS
Booster\footnote{\url{https://apps.fz-juelich.de/jsc/hps/juwels/jacamar.html}}
and these runners use Jacamar
CI\footnote{\url{https://ecp-ci.gitlab.io/index.html}}. The advantages of this
setup include its ease of use (as the batch scheduler details are hidden in the
JUBE script), no code duplication, and easy debugging. A disadvantage is the CB
run being marked as successful in the GitLab web interface as soon as JUBE
exits successfully, i.\,e., after successful batch scheduler job submission;
simulation failures are only discovered when the Python analyzer script is run
afterwards.
\nowidow[3]

\begin{figure}
	\input{gitlab-ci-libnegf-benchmarks-666b11bc.yml.tex.in}
	\caption{The CI configuration of libnegf-benchmarks. The command in line 11
	is a superficial check for the presence of the environment variables set
	below. If these are not exported by the callers, then the JUBE script will
	use default values.}
	\label{fig:libnegf-benchmarks-ci}
\end{figure}

\section{Case Studies: Diagnosing Software and Performance Issues}
\label{sec:case-studies}

\subsection{Diagnosing Rare Performance Issues}


With a benchmarking tool and the code forge in place, the libNEGF team
successfully tracked down a rare performance issue.

In September 2024 during a strong scaling run from eight to 192 nodes on JUWELS
Booster, the run on eight nodes unexpectedly timed out even with a generous
time limit 50\,\% above the estimated wall-clock time for the job. The problem
was immediately investigated with the aid of the artifacts saved by JUBE. After
a review no cause could be identified; human error was precluded as a cause.
Thereafter a GitLab issue was opened. The Jülich Supercomputing Centre
generates reports for every job; the tool generating these reports is called
\emph{LLview} \cite{FringsR2007}. These were attached to the issue and showed
the GPUs idling on all nodes except one. The code forge enabled a constructive
discussion with user's being able to add tables and images to their comments.
With JUBE in place, it was possible to check older runs for occurrences of this
problem and the problem had previously occurred without causing a time-out.
Regrettably the discussion did not lead to a quick resolution.

\begin{figure}
	\includegraphics[width=\textwidth]{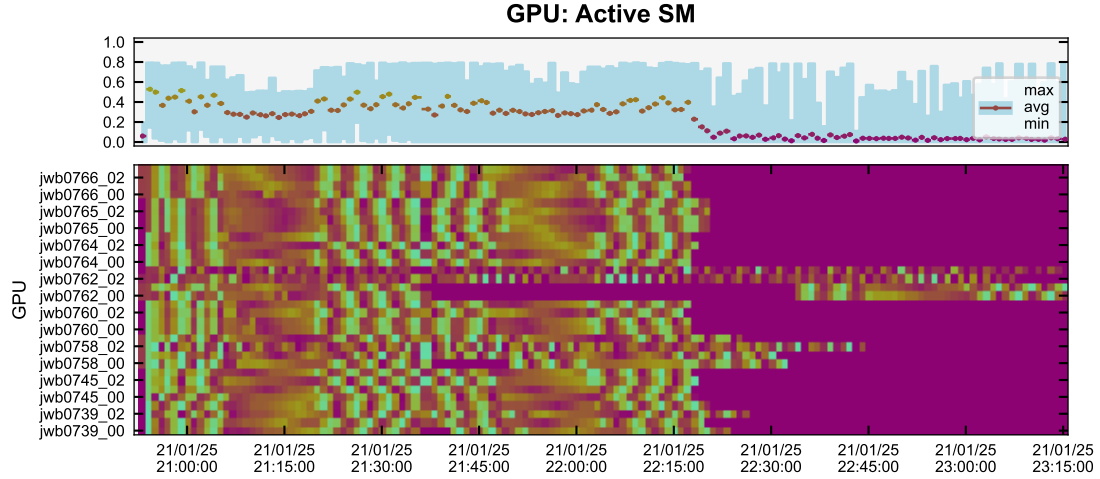}
	\caption{The GPU activity of a job on JUWELS Booster throughout its
	lifetime. The abscissa shows time and the ordinate the nodes. Idling is
	indicated by purple, high utilization by greenish colors. The plot was
	taken from the LLview job report.}
	\label{fig:libnegf-62-llview}
\end{figure}

Figure~\ref{fig:libnegf-62-llview} shows a part of an LLview job
report.\footnote{This figure was not taken from the job running on eight nodes
mentioned above but a different job exhibiting the same problem. This decision
was taken for expository reasons.} Note that most nodes start idling after
approximately 75~minutes. Even after accounting for load imbalance, the pattern
seen in the left portion (the part before 9:30\,pm) should be seen throughout
the lifetime of the job.
LLview job reports show memory consumption for each node over time and after
encountering the problem several times, a peculiar pattern began to emerge in
the memory consumption of concerned jobs: at the start of the job, the node
would already exhibit high memory consumption. This is unexpected because the
simulation's memory usage is initially low until all of the inputs are
consumed.

Given our understanding of the JUWELS Booster architecture, we were then able
to formulate an hypothesis about the cause of the slowdowns. Each JUWELS
Booster node possesses eight NUMA \cite[\S5.1]{CompArch} domains but only four
of those have a direct (and thereby fast) GPU connection. By default, libNEGF
is run with four MPI tasks per node with the tasks mapped to CPU cores whose
primary NUMA domain is one of the domains with direct GPU connection. Clearly,
if some foreign data is already occupying these NUMA domains, then libNEGF
would be slowed down. This theory puts the blame squarely on the HPC system. At
this point we contacted JSC support (Ticket \#1097823).

With JUBE in place, the JUBE data stored on JUWELS Booster, and the LLview job
reports, it was easy to convince the system administrators of our findings.
With their support the cause was tracked down to a Linux configuration quirk.
In general, the operating system caches filesystem data. Consider the
situtation where one NUMA domain is fully occupied by cached filesystem data
and the data of an application. When the application requests more memory, then
the operating system has to take a decision: either the cached data is freed to
satisfy the application's request or the application's request is fulfilled
with data from another NUMA domain. The best decision in this scenario is
application-dependent and for this reason the behavior can be
configured\footnote{See zone\_reclaim\_mode \cite{LinuxVM}}. Finally, the best
course of actions was found to be flushing the filesystem buffers before the
start of a new job. This resolved the issue.

\subsection{Tracking HPC Center Peformance}

The key advantage of CB is tracking application performance over time but the
performance is dependent on the underlying machine, too, and with CB one can
track the latter.

\begin{figure}[t]
	\includegraphics[width=\textwidth]{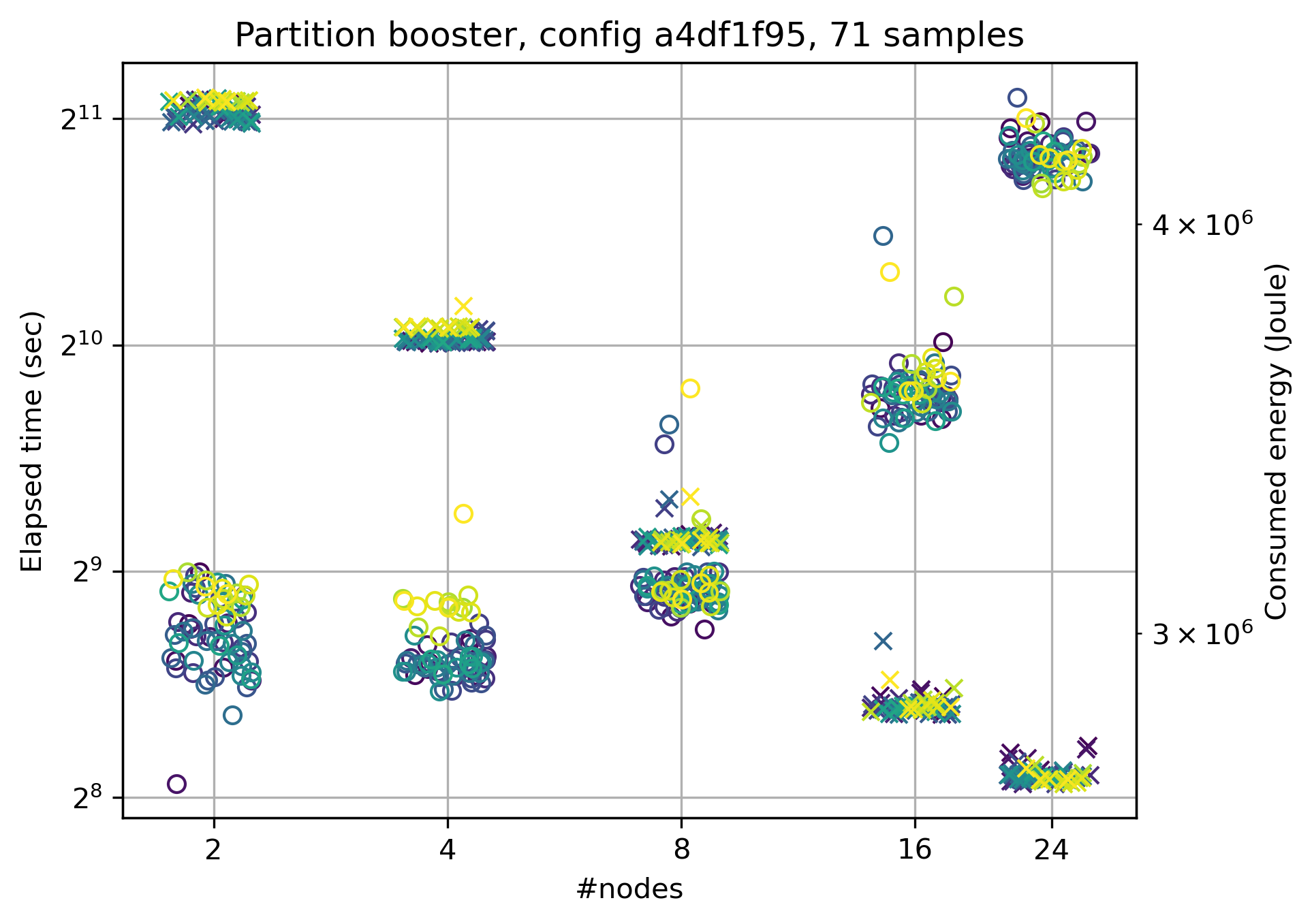}
	\caption{This plot shows the outcome of the strong scaling runs of the
	libNEGF CB setup. After a JUWELS Booster update on September 9, 2025,
	performance slightly decreased on two and four nodes.  Crosses indicate
	elapsed time (left axis), circles indicate energy consumption of the job
	(right axis), and the color indicates recency. To highlight the impact of
	the maintenance, jobs running after it are shown in bright colors and jobs
	running before it are shown in dark colors.}
	\label{fig:libnegf-cb}
\end{figure}

With CB in place for several weeks, we were able to identify a performance
degradation of the simulation code after JUWELS Booster maintenance without
\emph{any} changes in our code. Figure~\ref{fig:libnegf-cb} shows a plot of the
strong scaling runs of our CB setup with runs after the maintenance period in
bright colors. Observe that for two and four nodes, the run-time (indicated by
crosses) is higher after the update; on four nodes, the energy consumption
(indicated by circles) is also higher. For performance engineering purposes,
awareness of such events is important for the understanding of the performance
evolution of a software.

\subsection{Defect Chains hiding Misunderstood Mathematics}
\label{sec:broken-math}

In the previous sections, we mentioned several fixed libNEGF issues relating to
warnings and not checking return values. These issues turned out to be
connected and occluding a bigger issue.

There exists an iterative algorithm in libNEGF for dealing with boundary
conditions of the physical domain. Its input are two pairs of square matrices
$(S_1, H_1)$ and $(S_2, H_2)$. In a first step, complex scalars $\alpha_i$ and
linear combinations $A_i = \alpha_i S_i + H_i$ are computed, where $i = 1, 2$.
In a second step, the algorithm output is calculated from $A_1$, $A_2$ by an
iterative method.

This algorithm is implemented in CUDA and called directly from Fortran code.
CUDA code updates an iteration count in a memory location provided by the
caller of the CUDA code. Initially, the Fortran code was passing an iteration
counter by value to CUDA by mistake; ``by value'' implies that the Fortran
caller code never had access to the actual iteration count. The CUDA compiler
diagnosed this problem after enabling warnings. Once the iteration counter was
passed by reference, we started to check the counter at the call site after the
CUDA code had returned. We then found out that in one test of this algorithm,
the maximum iteration count was exceeded. Due to the simplicity of the test, we
arrived at the conclusion that the test was proper and that the algorithm
\emph{should} converge. After an investigation, we realized that our
comprehension of the algorithm was lacking. In the test $S_1$, $S_2$ were
chosen randomly but for convergence, $z S_2 + S_1 + z^{-1} S_2^*$ must be
Hermitian positive definite for all complex scalars $z$ with modulus one.
Luckily, for real-world inputs the matrices $S_i$ possess this property meaning
simulations were unaffected by this oversight.

In summary, two innocuous defects were concealing nonconvergence of an
algorithm with nonconvergence caused by our lack of understanding of certain
aspects of the algorithm. The defects were found using our suggested practices
and before our faulty assumptions affected production runs.

\section{Failure to Follow Best Practices}
\label{sec:no-best-practices}

\subsection{Standards Compliance}
\label{sec:standard}


libNEGF aims for portability across all major HPC systems and this includes
being able to compile libNEGF on these machines with the provided compilers.
Fortran is a standardized programming language but in practice, two problems
often arise: the code is not standards compliant or the compiler does not
implement the language standard in its entirety.
We found out that the libNEGF Fortran code does not compile with the Fujitsu
compiler (found on, e.\,g., Fugaku) nor with the Cray Fortran compiler (LUMI)
because libNEGF is not standards compliant. It just happened to be written in a
dialect compatible with the most frequently used compilers (GCC and Intel).
%
%
We are aware of one language feature in our code not supported by one of the
aforementioned compilers.

Our recommendation is to determine if portability is desired and if so,
compilers should be instructed to compile only standards-compliant code. This
change should happen as early as possible to avoid accumulating code that needs
to be fixed later. Our second recommendation is to check if the compilers on
target systems are fully supporting the Fortran subset in use by the project
because compiler support for a language standard may be extensive but not
complete.

\subsection{Lack of Defensive Programming}

libNEGF has been actively developed for more than a decade with its focus on
the prediction of material properties. As a side-effect, the software contains
no ``consistency'' checks, i.\,e., neither assertions about the program state
nor checks of the mathematical objects. This approach has drawbacks. First,
errors may not be detected at all unless a contributor with sufficient training
in the relevant physics takes a look at the simulation output. This is a
problem when the project's goal recently shifted to performance improvement and
a large number of simulation runs become a common event. Second, once an error
has been identified, the defect causing it has to be traced starting with a
complete simulation run.

Errors that went unnoticed for some time include broken GPU code and
nonsensical simulation output for unknown reasons.

We strongly urge other teams to have some kind of consistency checks to avoid
the aforementioned problems, \eg, assertions \cite[\S8.2]{McConnell2004},
backward error computations \cite[\S1.5]{Higham2002}, or checks of residuals.
These suggestions can be seen as defensive programming applied to scientific
software \cite[\S8]{McConnell2004}.

\subsection{Code formatting}


Proper code formatting helps contributors to understand code more quickly
\cite[\S31]{McConnell2004} and it avoids clutter in commits. With the latter we
mean that programmers may accidentally make semantically insignificant changes
to the code, \eg, adding empty lines, adding spaces at the end of a line, or
adding line breaks within existing code. These changes are picked up by the
revision control system (RCS). In severe cases, formatting differences in
semantically equivalent code may break the RCS' capabilities to merge different
code branches. This happened during libNEGF development where a fork of the
DFTB+ software had to be created to accommodate for newly added libNEGF GPU
code. After several months of development, an attempt a merging this branch
with the latest DFTB+ code in its main branch failed due to significant
formatting differences.

We recommend using an automatic code formatting tool. They are widely available
for almost all languages (even CMake) and literally work at the push of a
button. In our experience, the ease of use and the sensible layout generated
out of the box by these tools outweighs any concerns contributors may have with
regards to the style. They also save the programmer from memorizing large
documents for each language in use by a project (\eg, the Google C++ Style is
approximately 62~pages long\footnote{The Google C++ Style Guide was downloaded
on April 15, 2026 from \url{https://google.github.io/styleguide/cppguide.html}.
The document contains 31,074 words or approximately 62 pages at 500 words per
page.}). Code formatting should be the first thing to be checked in a CI
pipeline because it is fast and computationally cheap in comparison to other
typical operations.

\section{Conclusion}

Research software engineering techniques were applied when making libNEGF run
on massively parallel supercomputers. In Section~\ref{sec:software-engineering}
we presented our software engineering approach aimed at ensuring buildability
and defect-free code. The performance engineering approach detailed in
Section~\ref{sec:performance-engineering} simplifies benchmarking and
large-scale experiment runs for nonprogrammers. Our practices uncovered a
misunderstood mathematical model and performance changes induced by the HPC
cluster on which libNEGF ran (see Section~\ref{sec:case-studies}). Finally,
there are more common practices that should have been implemented by us as laid
out in Section~\ref{sec:no-best-practices}. By sharing our experiences we aim
to provide other RSE practitioners with data points to support their
decision-making. In particular, assuming code contains undetected defects
proved to be a useful engineering practice. In the case of libNEGF, untrapped
errors seem to be as common as for any other software written in an unsafe
language.

\begin{acknowledge}
	This project has received funding from the European High Performance
	Computing Joint Undertaking under grant agreement n°101144014.

	The authors gratefully acknowledge the Gauss Centre for Supercomputing e.V.
	(www.gauss-centre.eu) for funding this project by providing computing time
	through the John von Neumann Institute for Computing (NIC) on the GCS
	Supercomputer JUWELS\cite{Jsc2021} at Jülich Supercomputing Centre (JSC).
\end{acknowledge}

\begin{data}
	Figure~\ref{fig:libnegf-cb} can be reproduced from \cite{reproducer}. The
	LLview job report from which Figure~\ref{fig:libnegf-62-llview} was taken
	can be found in \cite{reproducer}, too.
\end{data}

\begin{aiuse}
	Successive drafts of this manuscript were repeatedly proofread with the
	help of Blablador \cite{Blablador}.
\end{aiuse}

\bibliographystyle{eceasst}
\bibliography{references.bib}

\end{document}